\documentclass[12pt]{article}
\usepackage{amsmath}
\usepackage{amssymb}
\renewcommand{\title}[1]{%
    \bigskip%
    \begin{center}%
    \Large\bf #1%
    \end{center}%
    \vskip .2in}

\renewcommand{\author}[1]{
   {\begin{center}
    #1
    \end{center}}}
\newcommand{\address}[1]{\vspace{-1.7em}\vspace{0pt}
    {\begin{center}
    \it #1
    \end{center}}}

\begin{document}
\title{\bf{ Energy momentum tensor of a non-minimally coupled scalar from the equivalence of the Einstein and Jordan frames}}

\author
{  
Sk. Moinuddin   $\,^{\rm a, b}$, 
Pradip Mukherjee  $\,^{\rm a,c}$,
Anirban Saha $\,^{\rm d,e}$,
Amit Singha Roy $\,^{\rm f,g}$ }

\address{$^{\rm a}$Department of Physics, Barasat Government College,Barasat, India}
\address{$^{\rm d}$Department of Physics, West Bengal State University, Barasat,  India}
\address{$^{\rm f}$Department of Physics, Cooch Behar Government Engineering College, Cooch Behar, India}
\address{$^{\rm b}$\tt  dantary95@gmail.com  }
\address{$^{\rm c}$\tt mukhpradip@gmail.com}
\address{$^{\rm e}$\tt  anirban@wbsu.ac.in }
\address{$^{\rm g}$\tt  singharoyamit@gmail.com }

\abstract{ Unlike the minimally coupled gravity theory where matter is coupled with gravity in such a manner so that one can differentiate the matter and gravity sector uniquely,  the non-minimally coupled theories (NMCT)  are distinguished by the intermingling of two. As a consequence of this the calculation of   the  energy momentum tensor (EMT) in  NMCT is beset with an arbitrariness. In this paper we provide an algorithm based on the well known equivalence between  Jordan frame and Einstein frame formulations which enables us to construct the EMT for NMCT in a unique way.}
%%Owing to it's non-minimal nature of coupling with gravity, the scalar field in a scalar-tensor theory is covariantly conserved with more than one choice of its energy-momentum tensor (EMT). Apparently this ambiguity in the choice of the symmetric EMT results from the way the conserved EMT is identified by different algebraic manipulations of the gravity field equation in such non-minimally coupled theories(NMC) . In this paper, we demonstrate that %using , one can arrive at these different choices of EMT, not by mere algebraic manipulation, but by demanding the equivalence of a NMC theory in Jordan frame with a minimally coupled theory in Einstein frame at the level of physical laws.

%%%%%%%%%%%%%%%%%%%%%%%%%%%%%%%%%%%%%%%%%%%%%%%%%%%%%%%%%%%%%%%%%%%%%%%%%%%%%%%%%%%%%%%%%%%%%%%%%%%%%%%%%%%%%%%%%%%%%%%%%%%%%%%%
\section{Introduction} 
%%%%%%%%%%%%%%%%%%%%%%%%%%%%%%%%%%%%%%%%%%%%%%%%%%%%%%%%%%%%%%%%%%%%%%%%%%%%%%%%%%%%%%%%%%%%%%%%%%%%%%%%%%%%%%%%%%%%%%%%%%%%%%%% 
 Einstein General Relativity (GR) is based on the action of the massive body on the space-time around. The latter again influences the masses self consistently. However the shear of gravitational field itself in the energy momentum tensor has been a controversial point right from the beginning. The principle of equivalence is considered in the weak form in GR\cite{Weinberg, Misner, Padmanabhan}. So there is no confusion in constructing the energy momentum tensor  (EMT) here. What one does is to place the matter in background gravity , watch the response of the system and the limit of this ratio of the changes $\delta{S}$ to the change in gravitational field $g_{\mu\nu}$ , in the limit of the $g_{\mu\nu}$ tend to ${\eta_{\mu\nu}}$ , gives us the EMT\footnote{The EMT in GR obtained as we have indicated has no reason to be equal to that   obtained by Noether's theorem\cite{Weinberg}.}. However in the scalar tensor theory initiated by the famous work of  Brans and Dicke \cite{nmo2},  a coupling was allowed between the two i.e matter and gravity. This makes the situation complicated because now one cannot be sure whether gravitation will act as a source for itself and if yes , how  ?  As far as we know, this issue is yet not resolved.  So interaction of matter with gravity is more difficult to study in such theories \cite{far1, far2, cap, cap2,cap3}.

The coupling of matter and gravity can be separated in GR . So it is called minimal coupling. With the advent of  experimental facilities in cosmology, it is known now that at the order of galactic cluster distance GR must be modified. But this modification would vanish in the solar system order\cite{Gonza, Iorio,Lin,Vfar,Ruggiero}. Of the post GR models where the coupling is not separable are called  non-minimally coupled theory(NMCT) \cite{farao,Carl,Maria,Hrycyna,Dav,Hertzberg, AASen,Obert}. It appears  that for  non-minimally coupled theory, it will be difficult to find an algorithm to construct the  EMT . This apprehension is corroborated by a plethora of papers on the subject \cite{NMC_quintescence_Phantom_crossing_2, Sotiriou,Cardone}. It has been attempted in the past to find  theoretically a method of construction based on field theory arguments and alongwith principle of equivalence \cite{pm} which was successful to reveal the inner link between the apparently different empirical methods of obtaining EMT.  But the algorithm \cite{pm} is so general that it did not serve the purpose of the practical cosmologist. However it gives a lesson that one has to use some extra general principle  for this type of construction.

 The non-minimal  type of coupling has gained popularity in recent past, because observational evidence in favour of the late time cosmic acceleration has opened up possibilities for such alternatives \cite{Bertolami,samy, NMC_our_paper}. These are the scalar-tensor theories\cite{Fujii, Dunn, Qui,Bartolo,pol,Troisi} and they are adopted as modified theories of gravity in numerous investigations \cite{cap,nmo3}.
 As the energy momentum tensor (EMT) of the scalar field in these scalar-tensor theories  can not be obtained using the standard definition of symmetric EMT\cite{Weinberg} , so different  prescription for writing the EMT are found in the literature . Though all these EMTs are covariantly conserved , the individual components of them are very different. However it is the individual component which appear in the equations of motion ,  so it is not clear how the different prescriptions lead to the same physical consequences. Hence an algorithm of obtaining the EMT for a non-minimally coupled theory is very much desirable.

 The usual algorithms used to construct the EMT in NMCT is to rearrange the gravitational field equation to mimic the Einstein field equations of GR \cite{NMC_quintescence_Phantom_crossing_2,DE,torre} , so that  an expression for the EMT may be identified .  While such a prescription is completely viable, this process of rearranging the gravitational field equation is not unique and leads to different expressions for the EMT of the same physical theory, as mentioned above. %It is rather curious to see the interrelation between the different EMTs . Whether such interrelation exists or not are rarely discussed. This means whether the dynamical  quantities like pressure , energy density and their ratio, the equation of state (EoS) parameter etc which are crucial elements  in the study of cosmology, have appropriate correspondence. to   obtained from these different choices of the EMT \cite{torre, FTEMT}, can describe the same physical system. 
In this paper, we will derive an algorithm  to construct an EMT exploiting the equivalance between  the conformally connected frames\cite{wald}.

  The new input in this paper is based on the formulation of the theory in two conformally connected frames. The first in which the physical model is defined and it is hence called physical frame or Jordan frame\cite{faraoni,faraoni1,Magnano,Liberati,Kubota,Catena,Salcedo}. The other frame is connected by the conformal transformation with the physical frame and it is called the Einstein frame. A notable property is the removal of the non-minimality in the Einstein frame. If other forms of matter is present then the non-minimality is shifted to that part. Now the other forms of matter are usually known matter( baryons, radiation and dark matter which can be relatively easy to tackle) and are not important for our general analysis. To simplify our analysis without missing any general connection we don't consider such terms.% We can derive the EMT theoretically  follows  from the fact that the NMCT can be formulated as a minimal theory by using a conformal transformation \cite{far1}\cite{far2} \footnote{In our analysis we are considering only the scalar field . But in general know forms of matter (Baryonic matter or radiation) may be present. Then the non-minimality is shifted on the matter Lagrangian.}.  This conformal connection is the extra input which we have mention earlier.The physical frame in which the NMCT is naturally formulated is called the Jordan frame and the frame in which the non-minimal coupling is removed is called the Einstein frame. Now it happens that the latter should be conformally connected with the Jordan frame  The connection between the analysis has been much debated at one time but now it is almost universally accepted that the two frames are definitely classically equivalent \cite{Postma, Sasaki,Nadeau,Aguilar}. Note that this is all we require.

 %Technically these are called the Jordan frame and the Einstein frame but one should note that no space time transformation connects these two.
% but can be arrived at starting from a unique description of the underlying physical system.
% \end{document}

%We enlist the established fact from two authorities in the field \cite{far2,wald} in section 2.

The organization of the paper is as follows. In section 2, we explain the process to be followed to obtain a symmetric, covariantly conserved EMT in the Jordan frame in a concise manner. 
 In section 3, we review how a suitable choice of conformal transformation converts  our scalar-tensor theory in the Jordan frame to a quintessence scalar field theory in the Einstein frame.  In section 4, we examine to what extent the equivalence of these two descriptions of the same physical system works. In section 5, we express the EMT obtained in the Einstein frame in terms of Jordan frame variables and also shading some light on it's nature.  In section 6,  we assume that the conservation law in the Einstein frame  implies the conservation law in the Jordan frame.  In view of the universal consensus about the equivalence of the Einstein and Jordan frames( at least in the classical level), these assumptions seems quite reasonable.  Using this equivalence, we have shown that %the different empirical forms assumed in the current work on scalar tensor theory are equivalent modulo covariantly conserved quantities. Thus the apparent empiricity of the EMT is solved.  Moreover 
 we can identify an appropriate EMT from these calculations.
%for all those choices of EMT that are conserved, thus establishing that the ambiguity in the EMT of the Jordan frame scalar field is only superficial and they represent the same physical system. 
We conclude in section 7 . The mostly positive signature of the metric is used throughout the paper.
 
 \section{Our approach}
% The introductory remarks  have made it amply clear that the purpose of the paper is to exploit the physical equivalence between the Jordan and Einstein frame formulations  of a NMCT for the construction of the EMT in the physical frame , so as to provide a theoretical stand point for the different empirical algorithm in-vogue .   The generic form of  NMCT is  given by
Our purpose is to provide an algorithm for construction of the EMT for NMCT which will depend on the canonical properties of the system and in no way  on any arbitrary assumption or physical intuition. The algorithm we propose is canonical in the sense that it is an action based method.  Remember that there exists no such theory till date . So our method if successful will lead to a novel algorithm for such an important physical variables as pressure, energy density etc.

A prototype non-minimally coupled theory will be assumed in a certain Friedmann Lemaître Robertson Walker (FLRW) spacetime which is written as 
\begin{eqnarray}
A_J
%&=&\int{{d^4}x}{\sqrt{-g}}\left[\frac{1}{16\pi G}\{1-\frac{\xi B(\pi)}{{(8\pi G)}^{-1}}\}R+\mathcal{L}_\pi\right]\nonumber\\
&=&\int{{d^4}x}{\sqrt{-g}}\left[ \frac{1}{2{\kappa}^2}D(\pi) R + \left\{-\frac{1}{2}g^{\mu\nu} {\nabla_\mu}\pi {\nabla_\nu}\pi-V(\pi) \right\} \right]
 \label{Action_Jordan}
\end{eqnarray}
Here 
\begin{equation}
D(\pi)=\left(1-\frac{\xi B(\pi)}{{(8\pi G)}^{-1}}\right)
\end{equation}

It characterizes the non-minimality with $B(\pi)$ being an arbitrary function of $\pi$ that can be tuned to give a class of non-minimally coupled theories. Units are chosen such that ${M_{pl}}^2=\frac{1}{8\pi G}=\frac{1}{\kappa^2}$ and for simplicity,{\bf{ we ignore all other matter fields}}.

 It has been proved quite generally that  a conformal transformation exists (\cite{Bonga})that maps the initial theory in FLRW to a flat Minkowski model. Our next task is to provide a conformal transformation which will map our FLRW manifold to a flat Minkowski manifold.

Let  $M$ be an $ n$-dimensional metric
 with Lorentzian  signature, and  $\Omega$ be
  a positive definite function then a transformation mapping it to the new space-time with metric,
  \begin {equation}
{\tilde{g}}_{\mu\nu } = \Omega^2 g_ {\mu\nu}
\label{ctm}
\end{equation} 
is called a conformal transformation.

\item A conformal transformation  in general is thus not equivalent to a diffeomorphism because of it's non-linearity. If the target and projected spaces have identical causal structure then they will be connected by a conformal transformation .

 Now we proceed with our construction. Since the target spacetime is a flat Minkowski spacetime( Einstein frame in our problem), it is easy to construct the EMT in this frame using the well known formula 
\begin{equation}
{\tilde{T}}_{\mu\nu}=-\frac{2}{\sqrt{-\tilde{g}}}\frac{\delta(\sqrt{-\tilde{g}}{L_\phi})}{\delta{\tilde{g}^{\mu\nu}}}\label{def_EMT}
\end{equation}
Automatically(\cite{Weinberg}) this EMT is divergence-less i.e. 
\begin{eqnarray}
\tilde{\nabla_\mu}{\tilde{T^{\mu\nu}}}= 0 \label{tal}
 \end{eqnarray}

We propose to transform the l.h.s. of (\ref{tal}) conformally to reach the initial state. This means in practice that the Einstein frame variables are substituted by Jordan frame variables.  We can do so because the conformal transformation is invertible. Note carefully, that the operator $\tilde{\nabla}$ when expressed in terms of Jordan frame variables following the given conformal connection, the result may not be of the same form as  (\ref{tal}).  However, from (\cite{wald}), we know that there exists conformal invariance of certain equations involving the metric. Just at this point we float our assumption that the form of the transformed equation (\ref{tal}) may (if necessary by utilizing the equation of motion of Jordan frame where ever required) be put as

\begin{eqnarray}
{\nabla_\mu}{{T^{\mu\nu}}_J}=0\label{ME1}
\end{eqnarray}
where all entities are Jordan frame variables.  ${{T^{\mu\nu}}_J}$ then can be thought of as the EMT of the NMCT under consideration in the Jordan frame.

\

 \section{From Jordan frame to Einstein frame}
%%%%%%%%%%%%%%%%%%%%%%%
In this section, we discuss the salient features of  our scalar-tensor theory as a non-minimally coupled scalar field interacting with gravity in the Jordan frame and briefly review how one can apply a suitable conformal transformation to convert it to a corresponding theory in the Einstein frame where the scalar field is minimally coupled to gravity. This will help fix our notations as well as summarize all the relevant transformation relations that we need for our purpose.
We start with the action for a scalar field $\pi$ non-minimally coupled to gravity in the Jordan frame given by (\ref{Action_Jordan}). The different symbols are explained therein. 
%\begin{eqnarray}
%A_J
%&=&\int{{d^4}x}{\sqrt{-g}}\left[\frac{1}{16\pi G}\{1-\frac{\xi B(\pi)}{{(8\pi G)}^{-1}}\}R+\mathcal{L}_\pi\right]\nonumber\\
%&=&\int{{d^4}x}{\sqrt{-g}}\left[ \frac{1}{2{\kappa}^2}D(\pi) R + \left\{-\frac{1}{2}g^{\mu\nu} {\nabla_\mu}\pi {\nabla_\nu}\pi-V(\pi) \right\} \right]
% \end{eqnarray}
%Here 
%\begin{equation}
%D(\pi)=\left(1-\frac{\xi B(\pi)}{{(8\pi G)}^{-1}}\right)
%\end{equation}

%characterizes the non-minimality with $B(\pi)$ being an arbitrary function of $\pi$ that can be tuned to give a class of non-minimally coupled theories.
% and 
%\begin{equation}
%{ \mathcal{L}}_\pi=-\frac{1}{2}g^{\mu\nu} {\nabla_\mu}\pi {\nabla_\nu}\pi-V(\pi)
%\end{equation}
%is the standard scalar field Lagrangean.
%Units are chosen such that ${M_{pl}}^2=\frac{1}{8\pi G}=\frac{1}{\kappa^2}$ and for simplicity, we ignore all other matter fields. Notice that due to the presence of $D(\pi)$ in it, the above action can not be separated into a dynamical action for gravity and that of a scalar field in curved spacetime like we usually do in minimally-coupled scalar field theories in curved spacetime. 
Now let us consider a conformal transformation given by (\ref{ctm})
%\begin{eqnarray}
%{\tilde{g}}_{\mu\nu}&=&{\Omega^2}g_{\mu\nu}
%\label{ct}
%\end{eqnarray}
which connects the metric $g_{\mu\nu}$ of the physical Jordan frame to a metric ${\tilde{g}}_{\mu\nu}$ on a different manifold. 
The determinants of these two matrices are related as %\footnote{Throughout the paper we use the symbol $\tilde{}$ to indicate quantities defined in the Einstein frame.} 
\begin{eqnarray}
\sqrt{-g}&=&\Omega^{-4}\sqrt{-{\tilde{g}}}
\label{ct_g}
\end{eqnarray}
and the affine connection in these two frames are related by
\begin{eqnarray}
\Gamma^{\alpha}{}_{\mu \nu} &=& \tilde{\Gamma}^{\alpha}{}_{\mu \nu}
-  \left[\tilde{\nabla}_{\nu}\left({\rm ln} \Omega \right) \delta^{\alpha}_{\mu}  + \tilde{\nabla}_{\mu}\left({\rm ln} \Omega \right) \delta^{\alpha}_{\nu}  - \tilde{\nabla}^{\alpha} \left({\rm ln} \Omega \right) \tilde{g}_{\mu \nu} \right] = \tilde{\Gamma}^{\alpha}{}_{\mu \nu} - \tilde{A}^{\alpha}{}_{\mu \nu}
\label{ct_gamma}
\end{eqnarray}
For future convenience let us note that although neither $\tilde{\Gamma}^{\alpha}{}_{\mu \nu}$ nor $\Gamma^{\alpha}{}_{\mu \nu}$ are tensor quantities in their respective frames, $\tilde{A}^{\alpha}{}_{\mu \nu}$ is a tensor quantity symmetric under $\mu \leftrightarrow \nu$. Moreover, its form is such that it can be readily written either in Jordan frame or in Einstein frame as per our convenience.
\begin{eqnarray}
\tilde{A}^{\alpha}{}_{\mu \nu} &=&  \left[\tilde{\nabla}_{\nu}\left({\rm ln} \Omega \right) \delta^{\alpha}_{\mu}  + \tilde{\nabla}_{\mu}\left({\rm ln} \Omega \right) \delta^{\alpha}_{\nu}  - \tilde{\nabla}^{\alpha} \left({\rm ln} \Omega \right) \tilde{g}_{\mu \nu} \right] \nonumber\\
&=& \left[\nabla_{\nu}\left({\rm ln} \Omega \right) \delta^{\alpha}_{\mu}  + \nabla_{\mu}\left({\rm ln} \Omega \right) \delta^{\alpha}_{\nu}  - \nabla^{\alpha} \left({\rm ln} \Omega \right) g_{\mu \nu} \right] = A^{\alpha}{}_{\mu \nu} 
\label{A}
\end{eqnarray}
Using the above relations (\ref{ctm}, \ref{ct_gamma}) we can further relate the curvature tensors defined in the two frames as
\begin{eqnarray}
R^{\alpha}{}_{\beta \mu \nu} = \tilde{R}^{\alpha}{}_{\beta \mu \nu} + 2 \tilde{A}^{\alpha}{}_{\beta \left[\mu; \nu \right]} + 2 \tilde{A}^{\alpha}{}_{\lambda \left[\mu \right.}  \tilde{A}^{\lambda}{}_{\left. \nu \right] \beta}
\label{ct_curvature}
\end{eqnarray}
and the corresponding Ricci scalars as
\begin{eqnarray}
R = \Omega^{2} \left[\tilde{R} + 6 \tilde{\Box} \left({\rm ln} \Omega \right) - 6 \tilde{\nabla}_{\mu}\left({\rm ln} \Omega \right)\tilde{\nabla}^{\mu}\left({\rm ln} \Omega \right)\right] 
\label{ct_R_scalar}
\end{eqnarray}
%as a candidate to transform the theory (\ref{Action_Jordan}) to a minimally coupled theory in the space connected to the physical Jordan space under the transformation (\ref{ct}).
Once we substitute equations (\ref{ctm}, \ref{ct_g}, \ref{ct_R_scalar}) the Jordan frame action (\ref{Action_Jordan}) takes the form
\begin{eqnarray}
A_J = \int d^{4}x {\sqrt{-\tilde{g}}} \left[ \frac{1}{2\kappa^2} \frac{D(\pi)}{\Omega^{2} } \{\tilde{R} + 6 \tilde{\Box} \left({\rm ln} \Omega \right) - 6 \tilde{\nabla}_{\mu}\left({\rm ln} \Omega \right)\tilde{\nabla}^{\mu}\left({\rm ln} \Omega \right)\} - \frac{1}{2\Omega^{2}} \tilde{g}^{\mu \nu}\tilde{\nabla}_{\mu} \pi \tilde{\nabla}_{\nu}\pi - \frac{V(\pi)}{\Omega^{4}} \right]
\label{nsi}
\end{eqnarray} 
Note that since $\pi$ is a scalar, $\nabla_{\mu} \pi = \tilde{\nabla}_{\mu} \pi$. Therefore the kinetic term for the scalar field $\pi$ can be written using the transformed metric.

Now a particular choice of the conformal transformation 
\begin{eqnarray}
D(\pi)=\Omega^2
\label{c_choice}
\end{eqnarray}
simplifies (\ref{nsi}) to
\begin{eqnarray}
A_J = \int{{d^4}x}{\sqrt{-\tilde{g}}} \left[ \frac{\tilde{R}}{2\kappa^2}+\frac{3}{\kappa^2}\tilde{\Box}\left( {{\rm ln}\Omega}\right)- \left( \frac{3}{\kappa^2} \tilde{\nabla}_{\mu}\left({\rm ln} \Omega \right)\tilde{\nabla}^{\mu}\left({\rm ln} \Omega \right) + \frac{1}{2\Omega^{2}} \tilde{\nabla}_{\mu}\pi \tilde{\nabla}^{\mu} \pi \right) - \frac{V(\pi)}{\Omega^{4}}\right]
\label{UE}
\end{eqnarray}
where the second term in (\ref{UE}) is a surface term since
\[ \tilde{\Box}\left( {{\rm ln}\Omega}\right) = \frac{1}{\sqrt{-\tilde{g}}} \partial_{\alpha}\left\{ \sqrt{-\tilde{g}} \, \tilde{\nabla}^{\alpha} \left( {{\rm ln}\Omega}\right)\right\}\]
Also due to the same choice of conformal transformation (\ref{c_choice}), 
\begin{eqnarray}
\tilde{\nabla}_{\mu}\left({\rm ln} \Omega \right) %= \left( \tilde{\nabla}_{\mu} \pi \right) \frac{d\left({\rm ln} \Omega \right)}{d \pi} 
=\frac{1}{2}  \frac{D^{\prime} }{D}\left( \tilde{\nabla}_{\mu} \pi \right)
\label{trivia}
\end{eqnarray}
with $\prime$ denoting derivative with respect to $\pi$. Using (\ref{trivia}) the third term in the parenthesis in equation (\ref{UE}) can be cast as the kinetic energy term of a new scalar field $\phi$
defined by
\begin{eqnarray}
{\left(\frac{d\phi}{d\pi}\right)}^2=\{\frac{3}{2\kappa^2}(\frac{D'}{D})^2+\frac{1}{D}\} = {\mathfrak{f}}^{2}(\pi)
\label{phi}
\end{eqnarray}
and finally we have our action (\ref{UE}) converted into
 \begin{equation}
A_E=\int{{d^4}x}{\sqrt{-\tilde{g}}}\left[\frac{\tilde{R}}{2\kappa^2}-\frac{1}{2}{\tilde{g}^{\mu\nu}} {{\tilde{\nabla}_\mu}\phi}{{\tilde{\nabla}_\nu}}\phi-U(\phi)\right] \label{Action_Einstein}
\end{equation}
in the transformed manifold where the newly defined scalar field $\phi$ behaves as a quintessence scalar field \cite{Harko, Caldwall, Steinhardt} with the self-interaction 
\begin{eqnarray}
U(\phi)=\frac{V(\pi)}{D^2}.
\label{pot_reln} 
\end{eqnarray}
%Remarkably, the action (\ref{Action_Einstein}) looks like a minimally coupled quintessence scalar field $\phi$ with potential $U(\phi)=\frac{V(\pi)}{D^2}$. 
The conformal transformation (\ref{ctm}) along with the choice (\ref{c_choice}) has mapped our non-minimally coupled scalar field theory in the Jordan frame to a minimally-coupled scalar field theory in this new frame with metric $\tilde{g}_{\mu\nu}$. This is referred to as the Einstein frame in standard literature \cite{DE}. %Since the Einstein frame is in a different manifold its scope is far reaching than the general coordinate transformations. 
The issue of the physical equivalence  of the Einstein frame to the Jordan frame is a sensitive one in the literature but as far as the classical aspects are concerned the two formulations can be safely assumed to describe the same physical reality. In the next section, let us critically pin-point to what extent this equivalence holds true.

%%%%%%%%%%%%%%%%%%%%%%%%%%%%%%%%%
\section{Equivalence of Jordan frame vis-\`{a}-vis Einstein frame}
%%%%%%%%%%%%%%%%%%%%%%%%%%%%%%%%%
Let us examine if the equivalence of Jordan frame vis-\`{a}-vis Einstein frame holds at the equation of motion level.
Because of the minimal coupling the action for the gravitational field and that of the scalar field are readily distinguishable in the Einstein frame action $A_{E}$ in (\ref{Action_Einstein}) 
\begin{eqnarray}
A_E&=&\int{{d^4}x}{\sqrt{-\tilde{g}}}\frac{\tilde{R}}{2\kappa^2} - \int{{d^4}x}{\sqrt{-\tilde{g}}}L_\phi \, %= S_G + S_\phi.
\label{rit}
\end{eqnarray}
So the standard definition of symmetric EMT (\ref{def_EMT}) for the scalar field $\phi$ 
readily applies here and yields
\begin{equation}
{\tilde{T}}_{\mu\nu} = \tilde{\nabla}_{\mu} \phi \tilde{\nabla}_{\nu}\phi - \tilde{g}_{\mu\nu} \{{\frac{1}{2}{\tilde{g}}^{\alpha\beta}} {{\tilde{\nabla}_\alpha}}\phi{{\tilde{\nabla}_\beta}}\phi+U(\phi)\}
\label{EFEMT}
\end{equation}
This symmetric EMT appears in the gravitational field equations 
\begin{eqnarray}
\tilde{G}_{\mu\nu}={\kappa^2}{{\tilde{T}}_{\mu\nu}}
\label{EF_EFT}
\end{eqnarray}
obtained by varying the action (\ref{Action_Einstein}) with respect to the Einstein frame metric $\tilde{g}_{\mu\nu}$. To obtain the equation of motion, i.e., the field equation for the scalar field $\phi$  we can either demand the covariant conservation of its EMT  
given by (\ref{tal})
%\begin{eqnarray}
%\tilde{\nabla}_{\alpha} \tilde{T}^{\alpha\beta} =0
%\label{cc}
%\end{eqnarray}
or vary the action (\ref{Action_Einstein}) with respect to $\phi$ to arrive at
\begin{eqnarray}
{\tilde{\Box}}{\phi} - \frac{d U}{d\phi}=0
\label{eqm_phi}
\end{eqnarray}
The equivalence of this Einstein frame description to its Jordan frame counterpart, on-shell, will be verified if starting from the field equation of $\phi$ (\ref{eqm_phi}) we can obtain the field equation for $\pi$, the corresponding scalar field in Jordan frame. To do this %, we reexpress equation (\ref{eqm_phi}) in terms of the Jordan frame variables, which is easy once 
we remember that both $\phi$ and $\pi$ are scalars, therefore their covariant derivatives are related by
 \begin{eqnarray}
{{\tilde{\nabla}}_\nu}\phi=\partial_{\nu} \phi = \left(\frac{d\phi}{d\pi}\right) \partial_{\nu}\pi = \left(\frac{d\phi}{d\pi}\right){\nabla_\nu}\pi={\mathfrak{f}}(\pi){\nabla_\nu}\pi
\label{d_equiv}
\end{eqnarray}
where in the last equality equation (\ref{phi}) is used. Using (\ref{d_equiv}) we can further relate their d'Alembertians as
\begin{eqnarray}
{\tilde{\Box}}\phi=\frac{({\mathfrak{f}}D)'}{D^2}\left({{\nabla_\alpha}\pi}{{\nabla^\alpha}\pi}\right)+\frac{{\mathfrak{f}}D}{D^2}{\Box}\pi
\label{box_reln}
\end{eqnarray} 
and from equation (\ref{pot_reln}) we calculate
\begin{eqnarray}
\frac{d U}{d\phi}=\frac{V'(\pi)}{{\mathfrak{f}}{D^2}}-\frac{2VD'}{{\mathfrak{f}}{D^3}}
\label{a}
\end{eqnarray}
Using these relations (\ref{box_reln}, \ref{a}) in (\ref{eqm_phi}) we get
\begin{eqnarray}
\Box \pi + \frac{\left({\mathfrak{f}}D\right)^{\prime}}{\left({\mathfrak{f}}D \right)} \left(\nabla_{\mu}\pi  \nabla^{\mu}\pi \right)  - \frac{V^{\prime}}{D {\mathfrak{f}}^{2}} - \frac{2V D^{\prime}}{D^{2}} = 0
\label{between}
\end{eqnarray}
Also note that $\pi$ is non-minimally coupled, so to get to its equation of motion, we need the gravitational field equation of the Jordan frame as well. The same can be obtained by varying the Jordan frame action (\ref{Action_Jordan}) w.r.t. $g_{\alpha\beta}$ that gives
\begin{eqnarray}
 D(\pi){G_{\alpha\beta}}={\kappa^2}\left[{\nabla_\alpha}\pi{\nabla_\beta}\pi-{g_{\alpha\beta}}\{\frac{1}{2}{g^{\mu\nu}}{\nabla_\mu}\pi{\nabla_\nu}\pi+V(\pi)\}\right]+\{{\nabla_\alpha}{\nabla_\beta}D(\pi)-{g_{\alpha\beta}}\Box D(\pi)\}
 \label{EFJF}
\end{eqnarray}
Using the trace of (\ref{EFJF})
\begin{eqnarray}
D(\pi)R%={\kappa^2}\left({{\nabla_\alpha}\pi}{{\nabla^\alpha}\pi}\right)+4{\kappa^2}V(\pi)+3\Box D(\pi) %\nonumber\\
=\kappa^{2}\left(1+\frac{3D''(\pi)}{\kappa^2}\right)\left({{\nabla_\alpha}\pi}{{\nabla^\alpha}\pi}\right)+3D'(\pi){\Box{\pi}}+4{\kappa^2}V(\pi)\label{JFT}
\end{eqnarray}
and re-expressing $({\mathfrak{f}}D)'$ using (\ref{phi}), as
\begin{eqnarray}
({\mathfrak{f}}D)'=\frac{(\frac{3D''}{\kappa^2}+1)\frac{D'}{2}}{({\mathfrak{f}}D)}\label{b}
\end{eqnarray} 
 in (\ref{between}) we can finally rewrite the $\phi$-field equation completely in terms of Jordan frame variables as
\begin{equation}
{\Box}\pi+D'\left[\frac{R}{2\kappa^2}\right]-V'(\pi)=0 \label{pi_fe}
\end{equation}
which is nothing but the equation of motion for the $\pi$- field in the Jordan frame, as can be verified by directly varying the Jordan frame action (\ref{Action_Jordan}) with respect to $\pi$. This  strengthens our conviction %that the criterion of equivalence is 
that physical behaviour of the system depicted in both the frames should map into each other exactly. However the geometric quantities of the two frames do not match, as can be seen from equations (\ref{ct_curvature}) and (\ref{ct_R_scalar}) that show both curvature tensors and Ricci scalars of the two frames differ by certain tensor quantities. This is not surprising since the basic geometry of the two frames encoded in their respective metrics are different (\ref{ctm}). In the next section, let us examine if physical parameters like energy density and pressure in the two frames adhere to this equivalence. 
%%%%%%%%%%%%%%%%%%%%%%%%%%%%%%%%%%%%
\section{Energy-momentum tensor in Einstein frame expressed in terms of Jordan frame variables }
%%%%%%%%%%%%%%%%%%%%%%%%%%%%%%%%%%%%
The parameters like energy density and pressure of the scalar field are encoded in its EMT. Owing to its minimal coupling to gravity the $\phi$ field in Einstein frame has an EMT (\ref{EFEMT}) that follows from an unambiguous definition (\ref{def_EMT}). So we start there and use equations (\ref{d_equiv}) and (\ref{pot_reln}) in (\ref{EFEMT}) to re-express it in terms of the $\pi$ field, its derivatives and its self-interaction $V\left( \pi \right)$. This gives 
\begin{eqnarray}
{{\tilde{T}}^{\alpha\beta}}&=&\frac{1}{D^2}\left[({{\nabla^\alpha}\pi}{{\nabla^\beta}\pi}){{\mathfrak{f}}^2}(\pi)-{{\mathfrak{f}}^2}(\pi)({{\nabla^\mu}\pi}{{\nabla_\mu}\pi})\frac{g^{\alpha\beta}}{2}-{g^{\alpha\beta}}\frac{V(\pi)}{D}\right]\nonumber\\
&\equiv& \Theta^{\alpha\beta}\left({\nabla^\mu}\pi,\pi \right)
\label{ecr1}
\end{eqnarray}      
We deliberately introduce a different symbol ${\Theta^{\alpha\beta}}({\nabla^\mu}\pi,\pi)$ here to signify that the right hand side is the Einstein frame EMT, but written in terms of Jordan frame scalar field and its derivatives. Direct calculation shows that $\Theta^{\alpha\beta}\left({\nabla^\mu}\pi,\pi \right)$ is not covariantly conserved in the Jordan frame and therefore can not pose as the EMT of the $\pi$ field. This is not entirely unexpected since the conformal transformation from Jordan to Einstein frames is not a change of variable but that of geometry. Therefore like geometric variables (e.g. curvature tensor) the physical observables in the two frames also do not map into each other.

 Before we proceed further let us  point out  the source of  ambiguity of the  EMT in any definition depending on the single action  (\ref{Action_Jordan}) -- it is due to the term $D(\pi)$. So to get an algorithm for EMT we must involve some extra input. In the present case it is the physical equivalence of   Einstein  and Jordan frame. How it is done will be described in the following.

\section{ Algorithm for the energy momentum tensor in the Jordan frame }
%%%%%%%%%%%%%%%%%%%%%%%%%%%%%%%%%%%
\noindent We start with the conservation law (\ref{tal}) for the scalar field $\phi$ in Einstein frame and try to re-express it in terms of Jordan frame variables. To this end, we first expand the covariant divergence of ${{\tilde{T}}^{\alpha\beta}}$
\begin{equation}
{{\tilde{\nabla}}_\alpha}{{\tilde{T}}^{\alpha\beta}}={\partial_\alpha}{{\tilde{T}}^{\alpha\beta}}+{\tilde{\Gamma}}^\alpha_{\alpha\lambda}{{\tilde{T}}^{\lambda\beta}}+{\tilde{\Gamma}}^\beta_{\alpha\lambda}{{\tilde{T}}^{\alpha\lambda}}
\end{equation}
and then using (\ref{ct_gamma}), (\ref{A}) and (\ref{ecr1}) write the expansion in terms of Jordan frame variables
\begin{equation}
{{\tilde{\nabla}}_\alpha}{{\tilde{T}}^{\alpha\beta}}={\partial_\alpha}{\Theta^{\alpha\beta}}+({\Gamma}^\alpha_{\alpha\lambda}+A^\alpha_{\alpha\lambda}){\Theta^{\lambda\beta}}+({\Gamma}^\beta_{\alpha\lambda}+A^\beta_{\alpha\lambda}){\Theta^{\alpha\lambda}}\label{A11}
\end{equation}
and finally combine suitable terms to express it as a sum of tensor quantities in the Jordan frame 
\begin{equation}
{{\tilde{\nabla}}_\alpha}{{\tilde{T}}^{\alpha\beta}}={\nabla_\alpha}{\Theta^{\alpha\beta}}+(A^\alpha_{\alpha\lambda}{\Theta^{\lambda\beta}}+A^\beta_{\alpha\lambda}{\Theta^{\alpha\lambda}})\label{A10}
\end{equation}
so that the conservation equation (\ref{tal}) of the Einstein frame, expressed in terms of Jordan frame variables, becomes
\begin{equation}
{\nabla_\alpha}{\Theta^{\alpha\beta}}+(A^\alpha_{\alpha\lambda}{\Theta^{\lambda\beta}}+A^\beta_{\alpha\lambda}{\Theta^{\alpha\lambda}}) =0 \label{A10a}
\end{equation}
Our aim is to show that we can extract  conserved EMTs from (\ref{A10a}) by simple algebraic manipulations and the equation of motion in Jordan frame. In other words we assume that the conformal transformations of the fields will allow us to write the expression 
\begin{equation}
{\nabla_\alpha}{\Theta^{\alpha\beta}}+(A^\alpha_{\alpha\lambda}{\Theta^{\lambda\beta}}+A^\beta_{\alpha\lambda}{\Theta^{\alpha\lambda}}) 
\end{equation}

 as a total divergent
 
 \begin{equation}
  {\nabla_\mu}{{T^{\mu\nu}}_J}
\end{equation}
  
where all entities of ${{T^{\mu\nu}}_J}$ are Jordan frame variables .   The research on non-minmal coupling is an old one with so many papers appearing in the field. The purpose is to explain the late time acceleration{\cite{Liddle}}. However, there is an opinion that the modifications of gravity can be absorbed in the equations of cosmology to project the theory as written in dark energy paradigm\cite{DE}. We do not subscribed to the view\cite{far1,far2} because there is deep physical considerations involved in the new modifications of gravity which cannot be just wiped away by a swing of hand. %So our view has originated from the references(\cite{far1,far2}).Most important purpose for which the present paper is mooted is discussed in the following.
%\subsection{The formula the Energy Momentum Tensor in the physical frame}
%Let us reiterate here that the definition of $\Theta^{\alpha\beta}$ in (\ref{ecr1}) means it {\it is} the EMT of the Einstein frame scalar $\phi$, but writen in terms of Jordan frame scalar field $\pi$ and its derivatives. We already know it has a non-zero divergence, so it is not the EM tensor, nevertheless it is a tensor quantity in Jordan frame.
%For our purpose we need to demonstrate that both versions (\ref{JFcc1}) and (\ref{JFcc2}) of the Jordan frame conservation law with their respective EMTs (\ref{ec1})  and (\ref{ec2}) emerge from (\ref{A10a}).
%Our final aim is to make it look like, one at a time. 

  So finally, we have reached a point from where the answer to the question posed in the very beginning of the paper is explicit. 
  
  %We follow a general treatment given in Wald( see equation (D.20) of appendix D of Wald).

Using (\ref{ecr1}) and the expression of $A^\alpha_{\mu \nu}$ from (\ref{A}), it is straightforward to compute
 \begin{equation}
\left( A^{\alpha}_{\alpha \lambda} \Theta^{\lambda \beta} + A^{\beta}_{\alpha \lambda} \Theta^{\alpha \lambda} \right) = \left\{2 \left( \nabla_{\lambda} \pi \nabla^{\lambda} \pi \right) \left( \frac{{\mathfrak{f}}}{D} \right)^{2} - \frac{V \left(\pi \right)}{D^{3}}\right\} \left( \frac{D'}{D} \right) \nabla^{\beta}\pi \label{inter}
\end{equation}
Similarly using (\ref{ecr1}) we can compute 
\begin{eqnarray}
{\nabla_\alpha}{{\Theta}^{\alpha\beta}} & = &\left\{\nabla_{\alpha}\left(\frac{  {\mathfrak{f}}  }{D} \right)^{2}\right\}\left\{{{\nabla^\alpha}\pi}{{\nabla^\beta}\pi} - \frac{1}{2}{g^{\alpha\beta}} \left( \nabla_{\lambda} \pi \nabla^{\lambda} \pi \right)\right\}  + \left(\frac{  {\mathfrak{f}}  }{D} \right)^{2} \nabla_{\alpha}\left\{{{\nabla^\alpha}\pi}{{\nabla^\beta}\pi} -  \right. \nonumber\\ &&  \left. \frac{1}{2}  {g^{\alpha\beta}}\left( \nabla_{\lambda} \pi \nabla^{\lambda} \pi \right)\right\}    - \nabla_{\alpha}\left(\frac{{g^{\alpha\beta}}{V(\pi)}}{D^3}\right)   
\label{simplify_1}
\end{eqnarray}

The second term of  equation (\ref{simplify_1}) can be written as 
\begin{eqnarray}
 \nabla_{\alpha}\left\{{{\nabla^\alpha}\pi}{{\nabla^\beta}\pi} - \frac{1}{2}{g^{\alpha\beta}} \left( \nabla_{\lambda} \pi \nabla^{\lambda} \pi \right)\right\} &=& \nabla_{\alpha}\left[{{\nabla^\alpha}\pi}{{\nabla^\beta}\pi} - \frac{1}{2} {g^{\alpha\beta}}\left( \nabla_{\lambda} \pi \nabla^{\lambda} \pi \right)        -{g^{\alpha\beta}}V(\pi)  \right.\nonumber\\
 &+&\left.   \frac{1}{\kappa^2}\{{{\nabla^\alpha}{\nabla^\beta}}D-{g^{\alpha\beta}}{\Box{D}}+(1-D){G^{\alpha\beta}}\} \right] \nonumber\\
 &&+ \nabla_{\alpha}\left[{g^{\alpha\beta}}V(\pi) - \frac{1}{\kappa^2}\{{{\nabla^\alpha}{\nabla^\beta}}D-{g^{\alpha\beta}}{\Box{D}}+\right. \nonumber\\ &&  \left.  (1-D){G^{\alpha\beta}}\}\right]
\label{simplify_2}
\end{eqnarray}

We can write the left hand side of (\ref{simplify_2}) as
\begin{eqnarray}
 \nabla_{\alpha}\left\{{{\nabla^\alpha}\pi}{{\nabla^\beta}\pi} - {g^{\alpha\beta}}\frac{1}{2} \left( \nabla_{\lambda} \pi \nabla^{\lambda} \pi \right)\right\} &=& \nabla_{\alpha}T^{\alpha\beta} + \nabla_{\alpha}\left[{g^{\alpha\beta}}V(\pi) -    \frac{1}{\kappa^2}\{{{\nabla^\alpha}{\nabla^\beta}}D  \right.\nonumber\\
 &-&\left. {g^{\alpha\beta}}{\Box{D}}+(1-D){G^{\alpha\beta}}\}\right]
\label{simplify_3}
\end{eqnarray}
where 
\begin{eqnarray}
T^{\alpha\beta}&=&\left[{{\nabla^\alpha}\pi}{{\nabla^\beta}\pi} - \frac{1}{2} {g^{\alpha\beta}}\left( \nabla_{\lambda} \pi \nabla^{\lambda} \pi \right)        -{g^{\alpha\beta}}V(\pi)\right.  \nonumber\\
 &+ & \left.  \frac{1}{\kappa^2}\{{{\nabla^\alpha}{\nabla^\beta}}D-{g^{\alpha\beta}}{\Box{D}}+(1-D){G^{\alpha\beta}}\}\right]\label{jemt}
 \end{eqnarray}

Thus equation (\ref{simplify_1}) becomes

\begin{eqnarray}
{\nabla_\alpha}{{\Theta}^{\alpha\beta}} & = & \left(\frac{  {\mathfrak{f}}  }{D} \right)^{2}\nabla_{\alpha}T^{\alpha\beta}+ \left\{\nabla_{\alpha}\left(\frac{  {\mathfrak{f}}  }{D} \right)^{2}\right\}\left\{{{\nabla^\alpha}\pi}{{\nabla^\beta}\pi} - \frac{1}{2}{g^{\alpha\beta}} \left( \nabla_{\lambda} \pi \nabla^{\lambda} \pi \right)\right\}  + \left(\frac{  {\mathfrak{f}}  }{D} \right)^{2}   \nabla_{\alpha}\left[{g^{\alpha\beta}}V(\pi)  \right.\nonumber\\
 &-&\left.   \frac{1}{\kappa^2}\{{{\nabla^\alpha}{\nabla^\beta}}D - {g^{\alpha\beta}}{\Box{D}}+(1-D){G^{\alpha\beta}}\}\right] - \nabla_{\alpha}\left(\frac{{g^{\alpha\beta}}{V(\pi)}}{D^3}\right)   
\label{simplify_11}
\end{eqnarray}

 The fact that $\nabla_{\alpha} V\left(\pi \right) = V^{\prime}\nabla_{\alpha} \pi $ and using the trace of the gravity field equation in Jordan frame (\ref{JFT}) and the equation of motion of the $\pi$ field (\ref{pi_fe}), we can further simplify (\ref{simplify_11}). The same is given by  
 
\begin{eqnarray}
{\nabla_\alpha}{{\Theta}^{\alpha\beta}} & = & \left(\frac{  {\mathfrak{f}}  }{D} \right)^{2}\nabla_{\alpha}T^{\alpha\beta}  -\left\{2 \left( \nabla_{\lambda} \pi \nabla^{\lambda} \pi \right) \left( \frac{{\mathfrak{f}}}{D} \right)^{2} - \frac{V \left(\pi \right)}{D^{3}}\right\} \left( \frac{D'}{D} \right) \nabla^{\beta}\pi  
\label{simplify_12}
\end{eqnarray} 
Putting the values of (\ref{inter}) and (\ref{simplify_12}) in (\ref{A10a}), one gets
 \begin{eqnarray}
\nabla_{\alpha} T^{\alpha\beta} =0
\end{eqnarray}

It is gratifying to observed that this is the sought for equation which we expected from our assumption. So we get the algorithm for finding EMT for any NMCT . 

For the convenience of the reader we now summarized our algorithm in the following
 \begin{enumerate}
\item  Let the action given in  Jordan frame be of the form (\ref{Action_Jordan}). Then the conformal transformation to the Einstein frame is to be found.
 
 \item  Using the Einstein frame action thus obtained and the formula (\ref{def_EMT}) the EMT in the Einstein frame (${\tilde{T^{\mu\nu}}}$) can easily be deduced. The corresponding conservation relation read as
 \begin{eqnarray}
\tilde{\nabla_\mu}{\tilde{T^{\mu\nu}}}= 0 \label{sk11}
 \end{eqnarray}
 
 Note carefully , that both the operator ($\tilde{\nabla_\mu}$) and EMT (${\tilde{T^{\mu\nu}}}$) are obtained as function of Einstein frame parameters.
 
 \item Now using the connection between  the Jordan frame and Einstein  frame we  get the form of  the conservation law (\ref{sk11}) as  
 
 \begin{equation}
{\nabla_\alpha}{\Theta^{\alpha\beta}}+(A^\alpha_{\alpha\lambda}{\Theta^{\lambda\beta}}+A^\beta_{\alpha\lambda}{\Theta^{\alpha\lambda}}) =0 \label{sk12}
\end{equation}
 Here all the variables are the function of Jordan frame parameters. 
 
 \item After the last step we are in the possession of an equation (\ref{sk12}) which is of the form $\chi=0$. Now by purely algebraic manipulations and the equation of motion of the Jordan frame we get the desired form (\ref{ME1}).

 \end{enumerate}

This EMT (\ref{jemt}) may be thought of as the conserved EMT in the Jordan frame i.e; $ {T^{\alpha\beta}}_J$ as mentioned in our approach (See below equation (\ref{tal})). Clearly in our algorithm there is no empirical division of the action or its arbitrary rearrangement .

% This is nothing but the EMT given in (\cite{torre}). So we believe that our algorithm will remove haziness present in the literature to obtain EMT of non-minimally coupled scalar in the Jordan frame.

%In a similar way, one can also show that \begin{eqnarray}\nabla_{\alpha} T_{[2]}^{\alpha\beta} =0\end{eqnarray}The emergence of the conservation law in the Jordan frame from that in the Einstein frame with both versions of the EMT of the $\pi$ field implies that both of them represent the same physical system. Despite their earlier appearance as the consequences of mere algebraic manipulations of the gravity field equation in Jordan frame, they actually have a more fundamental connection. 

%%%%%%%%%%%%%%%%%%%%%%%%%%%%%%%%%%%%%%%%%%%%%%%%%%%%%%%%%%%%%%%%%%%%%%%%%%%%%%%%%%%%%%%%%%%%%%%%%%%%%%%%%%%%%%%%%%%%%%%%%%%%%%%%

\section{Conclusion}
%%%%%%%%%%%%%%%%%%%%%%%%%%%%%%%%%%%%%%%%%%%%%%%%%%%%%%%%%%%%%%%%%%%%%%%%%%%%%%%%%%%%%%%%%%%%%%%%%%%%%%%%%%%%%%%%%%%%%%%%%%%%%%%%
The widespread use of scalar-tensor theories\cite{ samy, NMC_our_paper, Fujii,  Dunn, Qui} in cosmology demands a close examination of the ambiguity that is present in the energy momentum tensor (EMT) of the non-minimally coupled scalar. In NMCT , matter and gravity are so coupled that it is impossible to vary the matter action without varying gravity. Thus the form of EMT in this theory are empirically taken without any deep physical foundation.  The standard approach in the literature to circumvent this difficulty is to algebraically manipulate the gravity field equation so that the covariant conservation of the Einstein tensor can be used to identify the conserved EMT. However different manipulations lead to different forms of the EMT resulting in the said ambiguity. The importance of EMT in the cosmology can hardly be overemphasized. So such ambiguity in EMT coming from a mere algebraic rearrangement and not from some general physical principle is definitely not welcome. In this paper, we demonstrate how to extract a symmetric, covariantly conserved EMT from the conformal invariance \cite{wald,Postma,Sasaki,Nadeau,Aguilar, Bonga} of Jordan frame and Einstein frame  frames.  Interestingly, the non-minimally coupled theory in the Jordan frame emerges as a minimally coupled theory in Einstein frame description.  Though there remains some difference of opinion about the equivalence of these two formulations in the quantum mechanical level, it is universally accepted that classically the equivalence holds. 

In this paper, we employed this equivalence of the Jordan frame and Einstein frame formulation of the theory to explicitly demonstrate that physical features like time evolution of the fields and conservation laws in one frame implies that in the other. In the process we show that starting from the covariant conservation of the scalar field EMT in the Einstein frame one can arrive at the corresponding conservation law in the Jordan frame.%, but with different choices of the scalar field EMT \cite{ pm,torre} admissible and these different forms are precisely the ones that have been obtained in standard literature by algebraic manipulation of the gravity field equation. 
   
   So this paper serves a two-fold purpose, one is to examine the equivalence of the Einstein frame and Jordan frame descriptions \cite{faraoni,faraoni1,Postma,Carroll,capo}  of a class of scalar-tensor theories at the level of the conservation laws and the extraction of EMT of a NMCT in an unambiguous manner. Furthermore, our algorithm is class apart from others in the literature in the sense that we don't look for mathematical rearrangement of Einstein's gravity equation to obtain EMT of a NMCT rather we utilize equivalence of physical relations existing between frames to obtain a symmetric, covariantly conserved EMT.%another is to demonstrate that the various forms of the conserved EMT of the non-minimally coupled scalar are actually representative of the same underlying physical system. 

%%%%%%%%%%%%%%%%%%%%%%%%%%%%%%%%%%%%%%%%%%%%%%%%%%%%%%%%%%%%%%%%%%%%%%%%%%%%%%%%%%%%%%%%%%%%%%%%%%%%%%%%%%%%%%%%%%%%%%%%%%%%%%%%


\begin{thebibliography}{999}


\bibitem{Weinberg} S. Weinberg, ``Gravitation and Cosmology'' (Wiley, New York, 1972), p. 165.
\bibitem{Misner} Charles W. Misner,Kip S. Throne, J.A. Wheeler, ``Gravitation''( W. H. Freeman and Comapany, 1973)
\bibitem{Padmanabhan} T. Padmanabhan, ``Gravitation:Foundations and Frontiers''(Cambridge University Press)


\bibitem{nmo2} C. Brans and R. H. Dicke, Phys. Rev. 124 (1961) 925.



\bibitem{far1}
S. Capozziello, V. Faraoni ; {\it{``Beyond Einstein Gravity''}} ; 2011, Volume 170 ,ISBN : 978-94-007-0164-9.
\bibitem{far2}
 V. Faraoni ; {\it{Cosmology in Scalar-Tensor Gravity}} ; 2004, Volume 139 , ISBN : 978-90-481-6564-3 .
\bibitem{cap} S. Capozziello, M. D. Laurentis, Physics  Reports {\bf{509}} (2011), 167.
\bibitem{cap2} Salvatore Capozziello, Francisco S. N. Lobo, José P. Mimoso, Phys. Rev. D 91 : 124019 (2015)
\bibitem{cap3} Salvatore Capozziello, Mauro Francaviglia,Gen.Rel.Grav.40:357-420,2008






\bibitem{Gonza} P. A. González, Marco Olivares, Eleftherios Papantonopoulos, Yerko Vásq, The European Physical Journal C volume 80, Article number: 981 (2020) 
\bibitem{Iorio} Lorenzo Iorio, Matteo Luca Ruggiero, Scholarly Research Exchange, vol. 2008, article ID 968393
\bibitem{Lin} Wei-Ting Lin, Je-An Gu, Pisin Chen, 	arXiv:1009.3488
\bibitem{Vfar} Faraoni, V., 2006b, Phys. Rev. D74, 023529
\bibitem{Ruggiero} Ruggiero, M. L., and L. Iorio, 2007, JCAP 0701, 010






\bibitem{farao} Valerio Faraoni, Phys.Rev.{\bf D53} 6813-6821,1996.
\bibitem{Carl} Carlos R Fadragas, Genly Leon,  2014, Classical and Quantum Gravity 31(19).
\bibitem{Maria} Maria A. Skugoreva, Alexey V. Toporensky,Sergey Yu. Vernov, Phys. Rev. D 90, 064044 (2014).
\bibitem{Hrycyna} Orest Hrycyna, Marek Szydlowski,  JCAP11(2015)013.
\bibitem{Dav} Zahra Davari, Valerio Marra, Mohammad Malekjani, MNRAS 491 (2), 1920-1933 (2020).
\bibitem{Hertzberg} Mark P. Hertzberg, JHEP 1011:023,2010
\bibitem{AASen} Shruti Thakur, Anjan A Sen, T.R. Seshadri, 	Phys.Lett.B696:309-314,2011
\bibitem{Obert} Bertolami, O., 1987, Phys. Lett. B186, 161.



\bibitem{NMC_quintescence_Phantom_crossing_2} Fabio C. Carvalho, Alberto Saa, Phys.Rev. D70 (2004) 087302
\bibitem{Sotiriou} TP Sotiriou, V Faraoni, Rev. Mod. Phys. 82:451-497,2010 
\bibitem{Cardone} Capozziello, S., V. F. Cardone, and A. Troisi, 2005b, Phys.
Rev. D71, 043503.


\bibitem{pm}
Pradip Mukherjee, Amit Singha Roy, Anirban Saha, Mod.Phys.Lett.A {\bf {33}} (2018) 02, 1850010 .



%\bibitem{Moraes} P. H. R. S. Moraes , P. K. Sahoo, Eur. Phys. J. C (2017) 77: 480
%\bibitem{Ayuso} Ismael Ayuso, Jose Beltrán Jiménez, Álvaro de la Cruz-Dombriz, Phys. Rev. D 93, 089901 (2016)
\bibitem{Bertolami} O Bertolami, PJ Martins, Phys. Rev. D 61, 064007
\bibitem{samy} M. Sami, M. Shahalam, M. Skugoreva, A. Toporensky, PRD 86, 103532 (2012).
\bibitem{NMC_our_paper} Somnath Bhattacharya, Pradip Mukherjee, Amit Singha Roy, Anirban Saha, Eur.Phys.J. C78 (2018) no.3, 201.
%\bibitem{Fomin} I.V. Fomin, S. V. Chervon, A. V. Tsyganov, The European Physical Journal C volume 80, Article number: 350 (2020) %Why this paper...it does not mentions scalar-tensor theory even once!!!
\bibitem{Fujii} Yasunori Fujii and Kei-ichi Maeda, "The scalar-tensor theory of gravitation", Cambridge Monographs on Mathematical Physics (2003).
\bibitem{Dunn} K.A.Dunn, J. Math. Phys. 15, 2229 (1974).
\bibitem{Qui} Israel Quiros, Int.J.Mod.Phys.D28,No 7(2019)1930012.
\bibitem{Bartolo}  N. Bartolo and M. Pietroni, Phys. Rev. D61, 023518
(2000), hep-ph/9908521.
\bibitem{pol} G. Esposito-Farese, D. Polarski, Phys.Rev. {\bf D} 63 (2001) 063504.
\bibitem{Troisi}S. Capozziello, S. Nojiri, S.D. Odintsov, A. Troisi, Phys.Lett.B639:135-143,2006


\bibitem{nmo3}R. V. Wagoner, Phys. Rev. D 1 (1970) 3209.








\bibitem{DE} Luca Amendola, Shinji Tsujikawa, {\it Dark Energy: Theory and Observations}, Cambridge University Press, 2010.
\bibitem{torre} D.F. Torres, Phys.Rev.{\bf{D66}} (2002), 043522. 

\bibitem{wald}R.M.Wald ; {\it{General Relativity}} ; the university of Chicago ; 2006.





\bibitem{faraoni} Valerio Faraoni, Edgard Gunzig and Pasquale Nardone, Fund.Cosmic Phys.20:121,1999
\bibitem{faraoni1} Valerio Faraoni, Edgard Gunzig, Int.J.Theor.Phys. {\bf 38} (1999) 217-225.
\bibitem{Magnano} Magnano, G., and L. M. Sokolowski, 1994, Phys. Rev. D50,
5039.
\bibitem{Liberati}Sotiriou, T. P., V. Faraoni, and S. Liberati, Int.J.Mod.Phys.D17:399-423,2008
\bibitem{Kubota} T. Kubota, N. Misumi, W. Naylor and N. Okuda, JCAP 1202 (2012) 034
[arXiv:1112.5233 [gr-qc]]
\bibitem{Catena} R. Catena, M. Pietroni and L. Scarabello, Phys. Rev. D 76 (2007) 084039
[astro-ph/0604492]
\bibitem{Salcedo} I. Quiros, R. Garcia-Salcedo, J. E. M. Aguilar and T. Matos, Gen. Rel. Grav. 45 (2013)
489 [arXiv:1108.5857 [gr-qc]].

\bibitem{Liddle} An Introduction To Modern Cosmology, Andrew Liddle, John Wiley and Sons Inc,III River Street, Hoboken, NJ 07030, L'SA



\bibitem{Postma} Marieke Postma, Marco Volponi, Phys. Rev. D 90, 103516 (2014).

\bibitem{Sasaki} Guillem Dom`enech, Misao Sasaki, Int. J. Mod. Phys. D
25(13) (2016) 1645006 [1602.06332].
\bibitem{Nadeau} Faraoni, V., and S. Nadeau, 2007, Phys. Rev. D75, 023501.
\bibitem{Aguilar} I. Quiros, R. Garcia-Salcedo and J. E. M. Aguilar, arXiv:1108.2911 [gr-qc]

\bibitem{Bonga} Beatrice~Bonga, Kartik~Prabhu, Phys. Rev. D 102, 104043 (2020).

\bibitem {Carroll} S. Carroll, ``Spacetime and geometry: an introduction to general relativity''
\bibitem{capo} S Capozziello, R de Ritis, A A Marino, Class. Quantum Grav. {\bf 14} 3243 (1997).







 









 
%\bibitem{wald}
%R.M.Wald ; {\it{General Relativity}} ; the university of Chicago ; 2006 .



%\bibitem{Harko} Tiberiu Harko, Francisco S. N. Lobo, M. K. Mak, Eur.Phys.J.C74:2784,2014.
%\bibitem{Caldwall} R. R. Caldwell, Braz. J. Phys. 30 (2), June 2000.
%\bibitem{Steinhardt} Paul J. Steinhardt, Phil. Trans. R. Soc. Lond. A (2003) 361, 2497-2513.



%\bibitem{bacci} Francesca Perrotta, Carlo Baccigalupi, Phys.Rev. {\bf D65} (2002) 123505.
%\bibitem{H} Ingunn Kathrine Wehus, Finn Ravndal, Journal of Physics, Conference Series {\bf 66 }(2007) 012024.














\end{thebibliography}
\end{document}